\documentclass[]{aa}
\usepackage{graphics}
\begin{document}

   \thesaurus{06     
              (08.09.2 \object{WR 1};  
               08.13.2;  
               08.23.2)} 
   \title{An investigation of the large-scale \\variability of the
     apparently single Wolf-Rayet star \object{WR 1}}

   \subtitle{}

   \author{T. Morel\inst{1,}
   \thanks{\emph{Present address:} Astrophysics Group, Imperial College of Science, Technology and Medicine, Blackett Laboratory, Prince Consort Road, London, SW7 2BZ, UK; email: t.morel@ic.ac.uk}
          \and
           L. N. Georgiev\inst{2}
          \and
Y. Grosdidier\inst{1,}\inst{3}
          \and
N. St-Louis\inst{1}
          \and
T. Eversberg\inst{1,}
   \thanks{\emph{Present address:} Feinfocus Medizintechnik GmbH, Im Bahlbrink 11-13, 30827, Garbsen, Germany; email: t\_eversberg@feinfocus.com}
          \and
G. M. Hill\inst{4}}

   \offprints{T. Morel}
   \mail{t.morel@ic.ac.uk}

   \institute{D\'epartement de Physique, Universit\'e de Montr\'eal,
C. P. 6128, Succ. Centre-Ville, Montr\'eal, Qu\'ebec, Canada, H3C  3J7,
and Observatoire du Mont M\'egantic.\\
              email: morel, yves, stlouis, eversber@astro.umontreal.ca
        \and
             Instituto de Astronom\'{\i}a, UNAM, Apdo. Postal 70-264,
             M\'exico D. F. 04510, M\'exico.\\
              email:georgiev@astroscu.unam.mx
        \and
Observatoire Astronomique de Strasbourg, UMR 7550, 11 rue de l'Universit\'e, F-67000, Strasbourg,
             France.
             \and
             McDonald Observatory, HET, P. O. Box 1337, Fort Davis, TX.\\
             email:grant@astro.as.utexas.edu
}

   \date{Received ; accepted }

\titlerunning{Large-scale variability of the Wolf-Rayet star \object{WR 1}}
\authorrunning{Morel et al.}
   \maketitle

   \begin{abstract}

In recent years, much studies have focused on determining the origin of the large-scale line-profile and/or photometric patterns of variability displayed by some apparently single Wolf-Rayet stars, with  the existence of an
unseen (collapsed?) companion or of spatially extended wind structures as potential candidates. We present observations of \object{WR 1} which highlight the unusual character of the variations in this object. Our narrowband photometric
observations reveal a gradual increase of the stellar continuum flux
amounting to $\Delta v$ $\approx$ 0.09 mag
followed by a decline on about the same timescale (3-4 days). Only
marginal evidence for variability is found during the 11 following nights. Strong, daily line-profile variations are also
observed but they cannot be easily linked to the photometric variations. Similarly to the continuum flux variations, \emph{coherent}
time-dependent changes are observed in 1996 in the centroid, equivalent width, and
skewness of \ion{He}{ii} $\lambda$4686. 
Despite the generally coherent nature of the variations, we do not find evidence in our data for the periods claimed in previous studies. While the issue of a cyclical pattern of variability in \object{WR 1} is still controversial, it is clear that this object might constitute in the future a cornerstone for our understanding of the mechanisms leading to the formation of largely anisotropic outflows in Wolf-Rayet stars.

      \keywords{ Wolf-Rayet --
                 \object{WR 1} (HD 4004) --
                 mass loss
                                }
   \end{abstract}

%

\section{Introduction}
\rm It is now fairly well-established that \emph{apparently
single} Wolf-Rayet (WR)
stars may display two distinct (but probably non mutually exclusive) spectroscopic patterns of
variability: (a) small-scale emission features systematically moving
from the line center to the line wings on an hourly timescale (e.g., \cite{Lepinephd}); (b)
dramatically larger line-profile deformations  operating on a much longer
basis ($\sim$ days, e.g., \cite{Smithwillis}). Although the
first phenomenon, observed in most (if not all) WR stars, is believed to be the consequence of the fragmented,
possibly turbulent nature of the
outflow, the origin of the latter
type of variability is still very much elusive.  

Remarkable in this respect, is the existence of a well-established
(although strongly epoch-dependent) large-scale, \emph{cyclical} pattern of variability in the two apparently single WR stars \object{WR 6} ($\cal P$ =  3.763 $\pm$ 0.002 d; \cite{Firmani}) and \object{WR 134}
($\cal P$ = 2.27 $\pm$ 0.04 d; \cite{McCandliss94}; \cite{Morel98b}). A major observational effort has been directed on establishing the true nature of these peculiar
objects, i.e., whether this cyclical variability is induced by an orbiting unseen (collapsed?)
companion or by the rotational modulation of large-scale wind structures
(e.g., \cite{Vreux}; \cite{Morel98phd}; \cite{Harries}, and references therein). Although the
exact nature of these stars has yet to be unambiguously settled, these studies
reveal that rotational modulation constitutes an attractive alternative
to the binary hypothesis, especially
considering the recent recognition that some O stars (the progenitors of WR stars) might possess such azimuthally
structured 
 outflows (\cite{Fullerton}; \cite{Kaper}).

A prime target for further investigations is the seldom-studied, apparently single WN 5 star \object{WR 1} (HD
4004) that has recently been shown to present a spectral pattern of variability
very similar to that of \object{WR 6} (\cite{Niedzielski95}, 1996a, b;
Wessolowski \& Niedzielski 1996; \cite{Niedzielski99}). Strong line-profile
variability was observed, as well as apparently cyclical (according to $\cal P$ $\approx$ 2.667
days) variations in the EWs of \ion{He}{ii} $\lambda$4686 and \ion{He}{ii} $\lambda$5412 (\cite{Niedzielski96a}). The first claim
of periodic variability in \object{WR 1} with $\cal P$ $\approx$ 7.7 days was made by
   \cite{Lamontagnep} from an
analysis of the radial velocity
variations of \ion{He}{ii} $\lambda$4686.
The first photometric monitoring of this object
has shown  \object{WR 1} to be variable, with an indication of a 6.1 day period
(Moffat \& Shara 1986). Recently, \cite{Marchenko98a} discussed \emph{Hipparcos} broadband photometric data which revealed that \object{WR 1} also displays
relatively long-term photometric variations, with a marginal
evidence for a $\cal P$ = 11.68 $\pm$ 0.14 day periodicity.

As can be seen, controversy persits in the literature concerning the
possible cyclical nature of the variations in \object{WR 1}. We present in this
paper the results of spectroscopic and
photometric monitoring of \object{WR 1} carried out in 1995 and 1996
aiming at shedding some
light on this issue.
\section{Observations and Reduction Procedure}
\subsection{Photometry}
The photometric variability of \object{WR 1} has been investigated during the
interval 1996 September 18--October 5 by use of the single channel photometer \emph{Cuentapulsos} on the 0.84 m telescope of the
Observatorio Astron\'omico Nacional at San Pedro
M\'artir (Mexico). Two additional objects were
monitored during this observing run, namely, \object{WR 3} and \object{WR 153}. The nights were generally
clear. \object{WR 1} was observed through a
narrowband $v$ filter centered on 5140 \AA \ (FWHM = 90 \AA). This filter
samples a continuum-dominated region of the WR spectrum. We applied the
following sequence of 60 s integration through a 25\arcsec \ diaphragm:
sky, C2, C1, WR, C1, WR, C1, C2, sky. The same nearby comparison stars as used by
\cite{Moffatp} have been chosen. These comparison stars are similar in terms of their magnitude and colour to \object{WR 1}: $\Delta B$ (WR -- C1) = -- 0.17, $\Delta$[$B$ -- $V$] (WR -- C1) = -- 0.17, $\Delta B$ (WR -- C2) = -- 0.26, $\Delta$[$B$ -- $V$] (WR -- C2) = -- 0.19. An extinction coefficient $k_v$ = 0.20 was used throughout the data reduction. The scatter in the (C2 $-$ C1) data for the whole dataset amounts to $\sigma$ = 4.7 mmag. The differential magnitudes quoted in Table 1 are
averaged over two
consecutive cycles typically separated by about 20 minutes.
\begin{table*}
\caption{Journal of photometric observations}
\scriptsize
\hspace*{2cm} \begin{tabular}{cccc|cccc}
\hline \hline
HJD &&&&HJD&&&\\
(-- 2,449,000)&WR --  C1&WR --  C2&C2 --  C1 & (-- 2,449,000)&WR --  C1&WR --  C2&C2 --  C1\\\hline           
1345.721 & -- 0.043 & -- 0.114 & + 0.071 & 1356.693 & -- 0.054 & -- 0.121 & + 0.067\\
1347.702 & -- 0.074 & -- 0.134 & + 0.061 & 1356.773 & -- 0.057 & -- 0.128 & + 0.071\\
1347.731 & -- 0.063 & -- 0.121 & + 0.058 & 1356.847 & -- 0.065 & -- 0.130 & + 0.065\\
1347.760 & -- 0.064 & -- 0.137 & + 0.073 & 1356.902 & -- 0.054 & -- 0.127 & + 0.073\\
1347.846 & -- 0.065 & -- 0.134 & + 0.069 & 1356.984 & -- 0.054 & -- 0.130 & + 0.076\\
1347.882 & -- 0.066 & -- 0.141 & + 0.075 & 1357.656 & -- 0.064 & -- 0.139 & + 0.075\\
1348.807 & -- 0.095 & -- 0.163 & + 0.068 & 1357.723 & -- 0.064 & -- 0.129 & + 0.065\\
1348.831 & -- 0.105 & -- 0.173 & + 0.067 & 1357.781 & -- 0.055 & -- 0.121 & + 0.066\\
1348.857 & -- 0.094 & -- 0.170 & + 0.076 & 1357.847 & -- 0.058 & -- 0.127 & + 0.070\\
1348.886 & -- 0.103 & -- 0.161 & + 0.058 & 1357.914 & -- 0.057 & -- 0.129 & + 0.072\\
1349.678 & -- 0.126 & -- 0.197 & + 0.071 & 1358.735 & -- 0.055 & -- 0.126 & + 0.071\\
1349.761 & -- 0.119 & -- 0.185 & + 0.066 & 1358.773 & -- 0.059 & -- 0.125 & + 0.066\\
1349.854 & -- 0.129 & -- 0.195 & + 0.065 & 1358.829 & -- 0.048 & -- 0.109 & + 0.061\\
1349.933 & -- 0.127 & -- 0.193 & + 0.065 & 1358.881 & -- 0.061 & -- 0.123 & + 0.062\\
1349.991 & -- 0.130 & -- 0.194 & + 0.064 & 1358.943 & -- 0.051 & -- 0.117 & + 0.066\\
1350.873 & -- 0.101 & -- 0.164 & + 0.063 & 1358.992 & -- 0.058 & -- 0.120 & + 0.062\\
1350.947 & -- 0.104 & -- 0.167 & + 0.063 & 1359.642 & -- 0.051 & -- 0.118 & + 0.067\\
1351.006 & -- 0.100 & -- 0.164 & + 0.064 & 1359.702 & -- 0.040 & -- 0.112 & + 0.072\\
1352.851 & -- 0.059 & -- 0.123 & + 0.064 & 1359.767 & -- 0.043 & -- 0.115 & + 0.071\\
1352.913 & -- 0.053 & -- 0.126 & + 0.073 & 1359.832 & -- 0.049 & -- 0.116 & + 0.067\\
1352.953 & -- 0.054 & -- 0.119 & + 0.065 & 1359.909 & -- 0.051 & -- 0.117 & + 0.066\\
1352.964 & -- 0.052 & -- 0.118 & + 0.066 & 1359.958 & -- 0.050 & -- 0.122 & + 0.072\\
1353.000 & -- 0.058 & -- 0.125 & + 0.067 & 1360.004 & -- 0.059 & -- 0.122 & + 0.063\\
1353.649 & -- 0.062 & -- 0.125 & + 0.063 & 1360.632 & -- 0.063 & -- 0.133 & + 0.070\\
1353.747 & -- 0.058 & -- 0.126 & + 0.067 & 1360.697 & -- 0.046 & -- 0.114 & + 0.067\\
1353.836 & -- 0.048 & -- 0.115 & + 0.067 & 1360.761 & -- 0.059 & -- 0.126 & + 0.067\\
1353.936 & -- 0.049 & -- 0.115 & + 0.066 & 1360.842 & -- 0.057 & -- 0.116 & + 0.060\\
1354.013 & -- 0.060 & -- 0.126 & + 0.066 & 1360.912 & -- 0.061 & -- 0.123 & + 0.062\\
1354.671 & -- 0.051 & -- 0.124 & + 0.073 & 1361.637 & -- 0.060 & -- 0.122 & + 0.061\\
1354.754 & -- 0.057 & -- 0.133 & + 0.076 & 1361.694 & -- 0.060 & -- 0.132 & + 0.073\\
1354.810 & -- 0.048 & -- 0.124 & + 0.076 & 1361.768 & -- 0.061 & -- 0.121 & + 0.060\\
1354.875 & -- 0.047 & -- 0.113 & + 0.066 & 1361.851 & -- 0.065 & -- 0.124 & + 0.060\\
1354.948 & -- 0.062 & -- 0.129 & + 0.067 & 1361.952 & -- 0.065 & -- 0.131 & + 0.066\\
1355.002 & -- 0.054 & -- 0.121 & + 0.068 & 1361.997 & -- 0.058 & -- 0.126 & + 0.069\\
1355.668 & -- 0.044 & -- 0.113 & + 0.070 & 1362.625 & -- 0.053 & -- 0.118 & + 0.066\\
1355.735 & -- 0.056 & -- 0.124 & + 0.068 & 1362.720 & -- 0.060 & -- 0.124 & + 0.064\\
1355.792 & -- 0.058 & -- 0.121 & + 0.063 & 1362.846 & -- 0.052 & -- 0.121 & + 0.069\\
1355.860 & -- 0.060 & -- 0.123 & + 0.063 & 1362.910 & -- 0.055 & -- 0.123 & + 0.068\\
1355.912 & -- 0.050 & -- 0.121 & + 0.071 & 1362.983 & -- 0.053 & -- 0.123 & + 0.070\\
1355.993 & -- 0.064 & -- 0.116 & + 0.052 &         &        &        &  \\\hline
\end{tabular}
\end{table*}
\subsection{Spectroscopy}
\subsubsection{Long-slit and Reticon Spectra}
Long-slit spectra of \object{WR 1} have been obtained during various campaigns at the
Observatoire du Mont M\'egantic and Dominion Astrophysical Observatory
(Canada) in 1995 October and 1996 September. Reticon spectra were also obtained at DAO in 1996
November. The 1996 campaign at the
Observatoire du Mont M\'egantic was coordinated to support the
photometric campaign described above. Table 2 lists the mode of
observation, the dates of the spectroscopic observations, the interval of the observations
in heliocentric Julian dates, the
 observatory name, the number of CCD spectra obtained, the selected spectral domain, the reciprocal dispersion of the spectra, and the typical
signal-to-noise ratio (S/N) in the continuum.
\begin{table*}
\caption{Journal of spectroscopic observations}
\scriptsize
\begin{tabular}{llcccccc}
\hline \hline
\multicolumn{1}{c}{Mode of}  &   & HJD   & & Number & Spectral & Reciprocal Dispersion      &    \\
Observation & Date & (-- 2,449,000) & Observatory$^a$ & of Spectra & Coverage (\AA) &(\AA$\:$pix$^{-1}$)    & S/N \\\hline
Long-slit     & 95 Oct & 992-1005  & OMM, DAO & 7  & 3830-4300 & 1.6 & $\approx$ 160\\
    &         &             &          &  15  & 4360-5080 & 1.6 &
$\approx$ 170\\
    &         &             &          &  11  & 5050-5945 & 1.6 &
$\approx$ 290\\
Echelle & 96 Sep & 1342-1345 & SPM      & 72 & 3720-6900 & 0.16
(H$_\gamma$) --- 0.23 (H$_\alpha$) & $\approx$ 25 (H$_\gamma$) --- 50 (H$_\alpha$)\\
Long-slit     & 96 Sep & 1346-1352 & OMM      & 13   & 4400-5080 &
1.3 & $\approx$ 120\\
    & &             &          & 14  & 5050-6500 & 1.3 & $\approx$ 190\\
Reticon    & 96 Nov & 1386-1389 & DAO      & 12  & 5100-6110 & 0.5 &
$\approx$ 75\\\hline
\\\end{tabular}\\
$^a$ OMM: Observatoire du Mont
M\'egantic 1.6 m; DAO: Dominion Astrophysical Observatory 1.8 m; SPM: San Pedro
M\'artir Observatory 2.1 m.
\end{table*}
The spectra were reduced using the {\tt IRAF}\footnote{{\tt IRAF} is
distributed by the National Optical Astronomy Observatories, operated by
the Association of Universities for Research in Astronomy, Inc., under
cooperative agreement with the National Science Foundation.} data reduction
packages. The bias and sky subtraction, flat-field division,
removal of cosmic ray events, extraction of the
spectra, and wavelength calibration were carried out in
the usual way. Spectra of calibration lamps were taken immediately
before and after the stellar exposure. The stellar spectra were
subsequently continuum normalized by fitting
a low-order Legendre polynomial to carefully selected line-free
regions. In order to minimize the spurious velocity shifts induced by
an inevitably imperfect wavelength calibration, the spectra were
coaligned in velocity space by using the interstellar doublet \ion{Na}{i} $\lambda$$\lambda$5890, 5896 as fiducial marks. When
not available, the doublet \ion{Ca}{ii} $\lambda$$\lambda$3934, 3968 or the diffuse interstellar band at 4501
\AA \ were used. A trend for a systematic shift of the zero
point of the wavelength scale has been corrected by redshifting most of
the spectra by an average value of 35 km s$^{-1}$.
\subsubsection{Echelle Spectra}
 Echelle spectra have been obtained during the period 1996 September 16--19 with the Echelle
spectrograph \emph{Espresso} (\cite{Levine}) on the 2.1 m telescope of the San Pedro
M\'artir Observatory. The UCL
camera and a 1024 $\times$ 1024 coated CCD-Tek chip have been used. The selected grating (300 lines
mm$^{-1}$) yields a reciprocal dispersion of 0.16 and 0.23 \AA \
pixel$^{-1}$ at H$\gamma$ and H$\alpha$, respectively. The
spectra cover 27 orders and span the spectral range 3720-6900 \AA.

The reduction procedure (bias subtraction,
division by a normalized flat field, removing of scattered
light, extraction of the orders) was carried out
using the reduction tasks in the {\tt IRAF} package {\tt echelle}. Comparison spectra of Th--Ar lamps have been used for the wavelength
calibration. The typical accuracy of the wavelength calibration can be judged by
the dispersion in the heliocentric radial velocities of the
interstellar line \ion{Na}{i} $\lambda$5890: $\sigma$ $\approx$ 2 km s$^{-1}$. The
instrumental response has been removed by
fitting a low-order Legendre polynomial to the continuum sections of each order of an O-star spectrum, and then dividing the corresponding order of the \object{WR 1} spectra by this polynomial. The resultant, almost flat
orders were then combined using the {\tt IRAF} task {\tt scomb}. This procedure
proved to be satisfactory, except when the overlapping regions
fall in a spectral domain with very steep intensity gradients, as in the blue wing
of \ion{He}{ii} $\lambda$4686 (see Fig.\ref{f2}). The combined spectra were then subdivided into spectral
regions roughly corresponding to those of the long-slit spectra
discussed above (\S 2.2.1). For consistency purposes, these spectra have been continuum normalized by
fitting a low-order Legendre polynomial to the \emph{same}
continuum sections that selected for the long-slit spectra.
\section{Results}
\subsection{Photometric Variations}
\begin{figure}
\resizebox{\hsize}{!}{\includegraphics{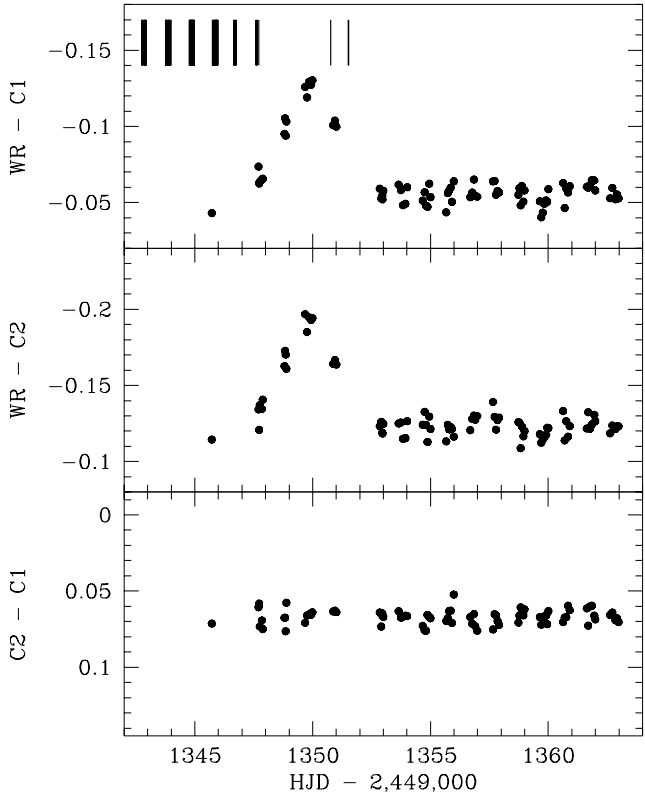}}
\caption{\emph{Two upper
panels}: differential $v$ magnitudes of \object{WR 1} in 1996 September relative to the two
comparison stars C1 and C2. \emph{Lower panel}: differential (C2 $-$ C1)
magnitudes. Dates when simultaneous spectroscopic observations were
performed are indicated by vertical tick lines in the uppermost panel.}
\label{f1}
\end{figure}

The light
curve of \object{WR 1} is plotted as a function of the heliocentric Julian date of
observation in Figure \ref{f1}. 
The main feature of this light curve is
the gradual increase of the stellar continuum flux amounting to $\Delta v$ $\approx$ 0.09 mag
beginning at HJD 2,450,346, followed by its decline on about the same
timescale after HJD 2,450,350. During the last 11 nights (after HJD 2,450,353), only marginal evidence for variability is found, with
$\sigma$(WR $-$ C)/$\sigma$(C2 $-$ C1) $\approx$ 1.3. This can be compared
with a value
of 1.9 derived by \cite{Moffatp} on the basis of 14 nights of broadband $B$ observations in
1984.
Noticeable is the striking
lack of recurrence in the continuum flux data. A periodicity
search in the (WR $-$ C1) and (WR $-$ C2) data acquired after HJD 2,450,353
yields no evidence for any periodic signals in the range 6--8 days that would
possibly account for the variations observed before this
date. The highest
peaks in the power spectra of these two datasets (after correction by
the CLEAN algorithm; see \cite{Roberts})
appear (with considerable
uncertainty) for periods of about 4 days. However, the significance of
these periodic signals is very low and
they must thus be
regarded as spurious. The periods proposed by Niedzielski (1996a; 2.667 days) and  Marchenko et al. (1998a; 11.68 days) are inconsistent with the global light-curve morphology. On the other hand, the periods claimed by Lamontagne (1983; 7.7 days) and Moffat \& Shara (1986; 6.1 days) are only consistent with part of the data (before HJD 2,450,353). We will come back to this point later (\S 4).

\subsection{An Overview of the Line-profile Variations}
\begin{figure}
\resizebox{\hsize}{!}{\includegraphics{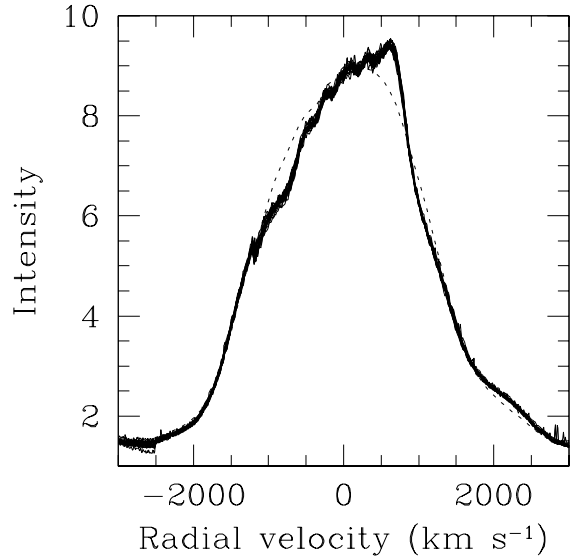}}
\caption{Superposition of the 21 Echelle spectra
acquired on 1996 September
18 (around HJD 2,450,344.873) for the spectral range encompassing \ion{He}{ii} $\lambda$4686. These spectra have been obtained on a total time interval of about 5.5
hr. For
comparison purposes, the
dashed line shows the mean spectrum of the 1996 September dataset. The discontinuities
observed at about $-$ 2500 km s$^{-1}$ and $-$ 1200 km s$^{-1}$ \ are artefacts of an
imperfect connection of the orders. The
projected velocities, as everywhere in this
paper, are heliocentric and refer to the line laboratory rest wavelength.}
\label{f2}
\end{figure}

\begin{figure*}
\resizebox{\hsize}{!}{\includegraphics{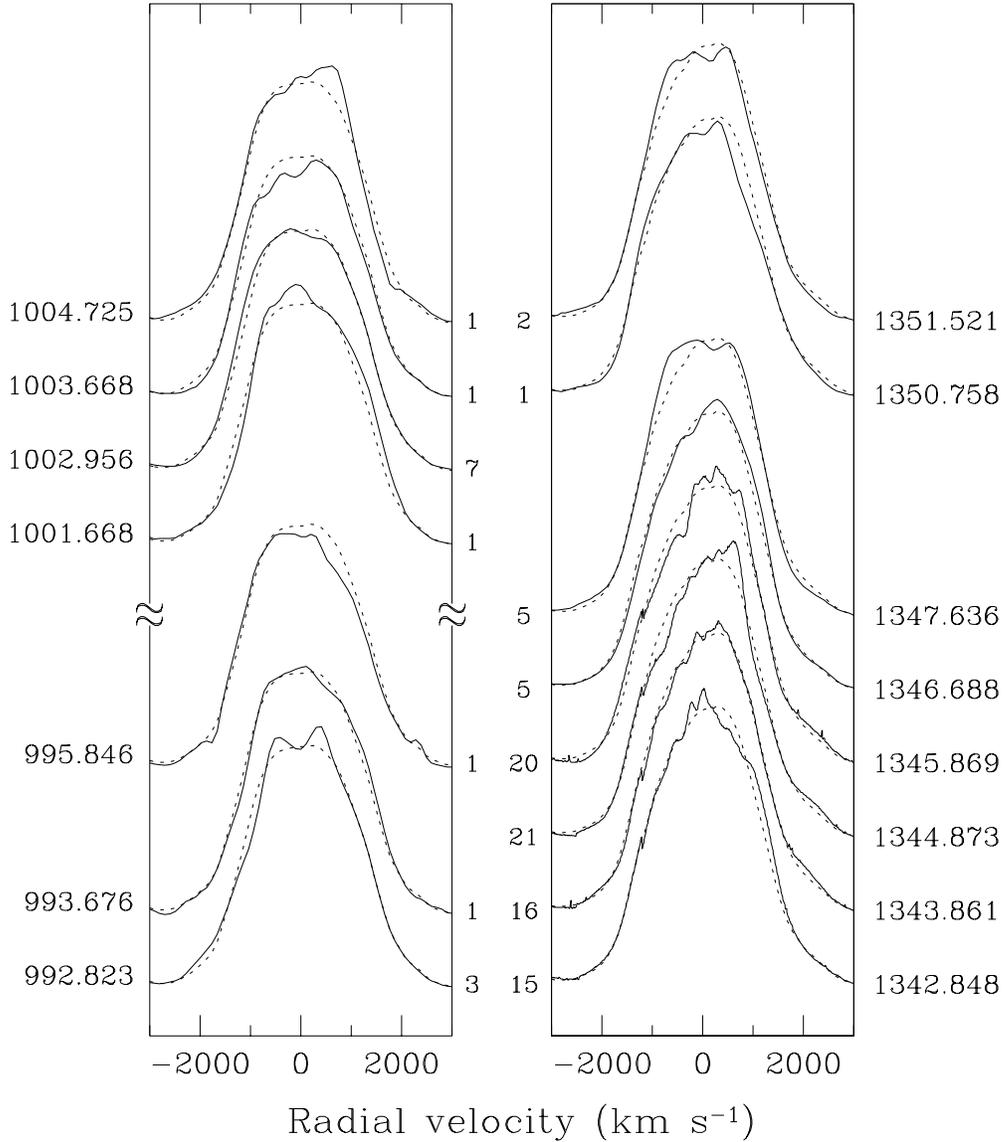}}
\caption{Montage of the continuum-normalized, nightly
mean spectra obtained in 1995 October (\emph{left panel}) and 1996 September (\emph{right panel}) for
the spectral range encompassing \ion{He}{ii}
$\lambda$4686 (the number of individual spectra
averaged to form the corresponding nightly mean is indicated between
the two panels). The mean heliocentric Julian date of the observations
(-- 2,449,000) is indicated on both sides of
the panels. For comparison purposes, the unweighted means for 1995 October (7 spectra) and 1996 September
(8 spectra) are overplotted as dashed
lines. The spectra of consecutive nights are offset by 2.5 units of continuum in the intensity
scale.}
\label{f3}
\end{figure*}

In the following, we will generally illustrate the line-profile variations of \object{WR 1} by means of the strong \ion{He}{ii} $\lambda$4686 line as its variations are qualitatively similar to those affecting the other \ion{He}{ii} features (see \S 3.5; \cite{Niedzielski96a}). One of the most outstanding features of the line-profile variability is the relatively low level of variability observed within one night.\footnote{Note that although probably present in \object{WR 1}, the small-scale emission-excess features
travelling on an hourly timescale on the
top of the line profiles (probably induced in WR stars by
outwardly moving shocked and/or turbulent material; \cite{Lepine})
cannot be studied here, owing to the insufficient S/N achieved for
the Echelle spectra.} Such a gradual pattern of variability has been observed whenever several spectra have been obtained during a single night of observation. This phenomenon is illustrated in Figure \ref{f2} where we show a superposition of the 21 \ion{He}{ii} $\lambda$4686 line profiles obtained on the night of 1996 September
18 (around HJD 2,450,344.873). 
In sharp contrast, however, noticeable night-to-night variations are observed. This is illustrated in Figure \ref{f3} which shows a montage of the nightly mean spectra obtained in
1995 October (\emph{left panel}) and 1996 September (\emph{right panel}) for
the spectral region encompassing \ion{He}{ii} $\lambda$4686 (it has to be kept in mind that because of
the paucity of the photometric data, the strength of the spectral lines in these continuum-normalized
spectra has not been corrected for the varying continuum flux level).
\subsection{Temporal Variance Spectrum Analysis}
\begin{figure*}
\resizebox{\hsize}{!}{\includegraphics{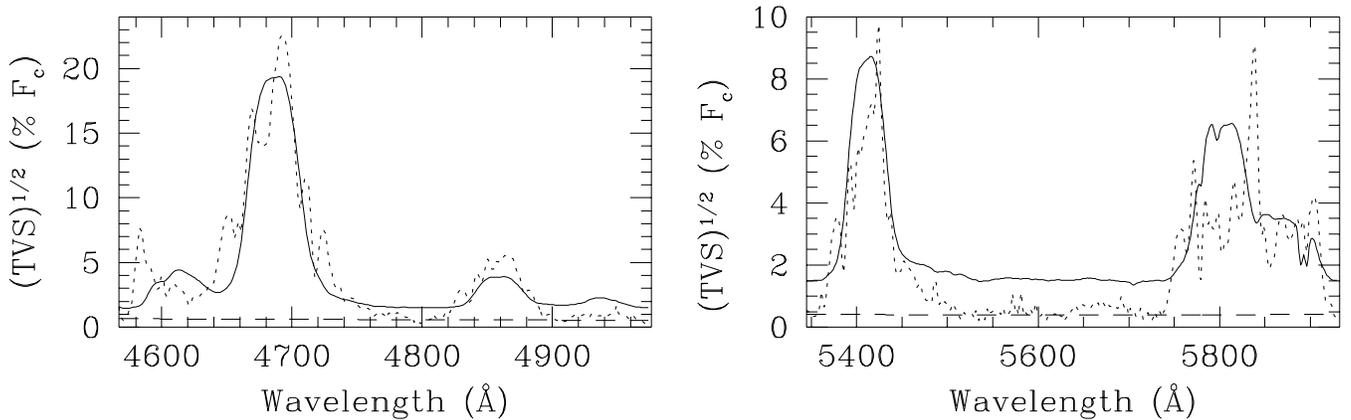}}
\caption{Square root of the Temporal Variance Spectrum (\emph{short-dashed line}), along with the threshold indicating the 99.0 \%
confidence level for significant variability (\emph{long-dashed line}). The mean of the spectra used in the calculations is overplotted in arbitrary units (\emph{solid line}).}
\label{f4}
\end{figure*}

The Temporal Variance Spectrum (TVS; \cite{Fullertontvs}) has been used to assess the level of spectral variability as a function of wavelength. The (square root of the) TVS, giving the typical ``size'' of the
deviations from a template-weighted mean spectrum (expressed in percentage of the continuum
flux) is shown, along with the 99.0 \% confidence level for variability,
in Figure \ref{f4}. 
As can be seen, highly significant variability affects the emission lines, with the notable exception of \ion{N}{v} $\lambda$ 4945 for which only little evidence for variability is found. Many peaks (i.e., locations of ``preferential'' variability) can be distinguished in the TVS. The locations of these peaks (in terms of the projected velocity referred to the laboratory rest wavelength of the spectral feature in question) are quoted in Table 3. This investigation shows that: (a) the variability often extends to velocities comparable to the wind terminal velocity ($v_{\infty}$ $\approx$ 2135 km s$^{-1}$; \cite{Rocho}), e.g., $v$ $\approx$ -- 1790 km s$^{-1}$ for the relatively unblended line \ion{He}{ii} $\lambda$4859; (b) the TVS structure --- made up of several subpeaks --- presents some similarities for the various \ion{He}{ii} features. However, the velocities quoted in Table 3 are certainly entached of considerable uncertainties (e.g., because of blending of TVS subpeaks) making difficult a clear statement as to whether the variability takes place at the same characteristic velocities in different lines.
\begin{table}
\caption{Projected velocities of the TVS subpeaks observed in \ion{He}{ii} $\lambda$4686, \ion{He}{ii} $\lambda$4859, and \ion{He}{ii} $\lambda$5412.$^a$}
\hspace*{1cm} \begin{tabular}{ccc}
\hline  \hline  
\ion{He}{ii} $\lambda$4686 & \ion{He}{ii} $\lambda$4859 & \ion{He}{ii} $\lambda$5412\\\hline
-- 2260 & -- 1790 & -- 1880\\
-- 1060 & -- 1060 & -- 1040\\
        & -- 550  &        \\
+ 430   & + 290   & + 690  \\
+ 1560  & + 1390  & + 1430 \\
+ 2460  &         &        \\\hline
\\\end{tabular}\\
$^a$ In km s$^{-1}$ and referred to the line laboratory rest
wavelength.
\end{table}
\subsection{The P Cygni Profile Variability}
\begin{figure}
\resizebox{\hsize}{!}{\includegraphics{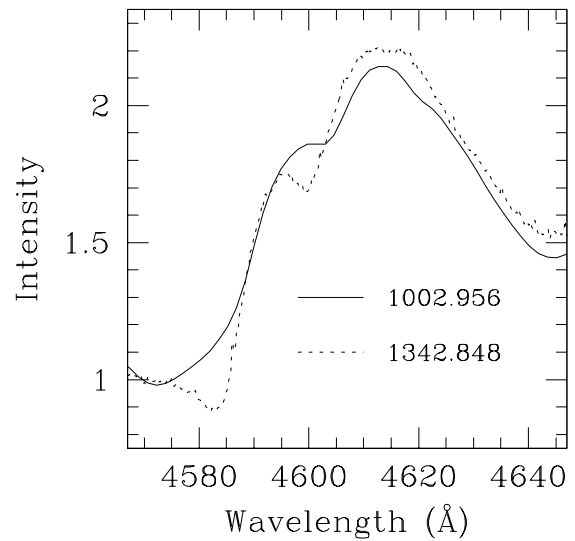}}
\caption{Superposition for
the spectral range encompassing \ion{N}{v} $\lambda$$\lambda$4604, 4620 of 
 the nightly means of 1995 October 12 (\emph{solid line}) and 1996 September 16
(\emph{short-dashed line}). The mean heliocentric Julian dates of the
observations (-- 2,449,000) are indicated.}
\label{f5}
\end{figure}

The TVS analysis also reveals substantial variability at the location of the P Cygni absorption component of \ion{N}{v}
$\lambda$4604 (Fig.\ref{f4}). This mainly results from the transition of the \ion{N}{v}
$\lambda$4604 feature from a pure emission line-profile in 1995 October to a P Cygni line-profile in 1996 September (Fig.\ref{f5}). No clear
variations in the strength of this absorption trough on a daily
timescale are observed.

\begin{figure*}
\resizebox{\hsize}{!}{\includegraphics{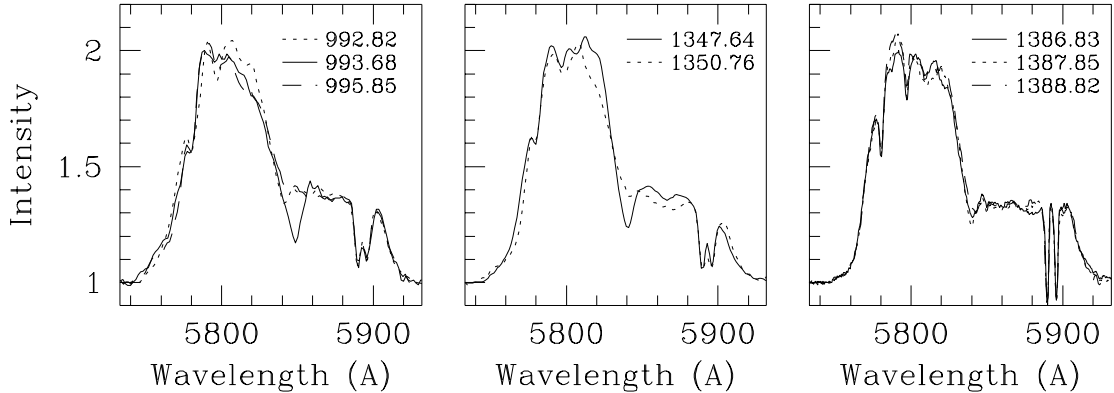}}
\caption{Superposition for
the spectral range encompassing \ion{C}{iv} $\lambda$5806 and \ion{He}{i} $\lambda$5876 of nightly means observed in 1995 October (\emph{left panel}), 1996 September (\emph{middle
panel}), and 1996 November (\emph{right
panel}). The mean heliocentric Julian dates of the observations
(-- 2,449,000) are indicated in each panel.}
\label{f6}
\end{figure*}

An enhanced peak in the TVS can also be found at the location where the carbon triplet \ion{C}{iv} $\lambda$5806 and
\ion{He}{i} $\lambda$5876 merge (Fig.\ref{f4}). The large daily changes affecting \ion{C}{iv} $\lambda$5806 and/or
\ion{He}{i} $\lambda$5876 in October 1995 and September 1996 are illustrated in Figure \ref{f6} (this phenomenon is not observed in 1996 November). The high level of variability observed at this particular location is likely due to the superposition of two distinct type of variability: (a) red-wing variability of \ion{C}{iv} $\lambda$5806 as observed in other spectral features (e.g., \ion{He}{ii} $\lambda$4686; Fig.\ref{f4}); (b) variations in the strength of the P Cygni absorption component of  \ion{He}{i} $\lambda$5876.\footnote{Such P Cygni absorption troughs of \ion{He}{i} $\lambda$5876 are commonly (and more clearly) observed in other WN stars (e.g., \cite{Robert}).} The latter interpretation is supported by the fact that the projected velocity of the peak in the TVS ($v$ $\approx$ -- 1980 km s$^{-1}$, referred to the \ion{He}{i} rest laboratory wavelength) matches fairly well the wind terminal velocity ($v_{\infty}$ $\approx$ 2135 km s$^{-1}$). Considering the relatively modest level of variability affecting the red-wing part of the line profiles in \object{WR 1} (Fig.\ref{f4}), it is likely that changes in the
strength of the P Cygni absorption component of the
\ion{He}{i} feature \emph{mostly} contribute to the observed changes.
\subsection{Correlated Line-profile Variations in Different Lines?}
\begin{figure*}
\resizebox{\hsize}{!}{\includegraphics{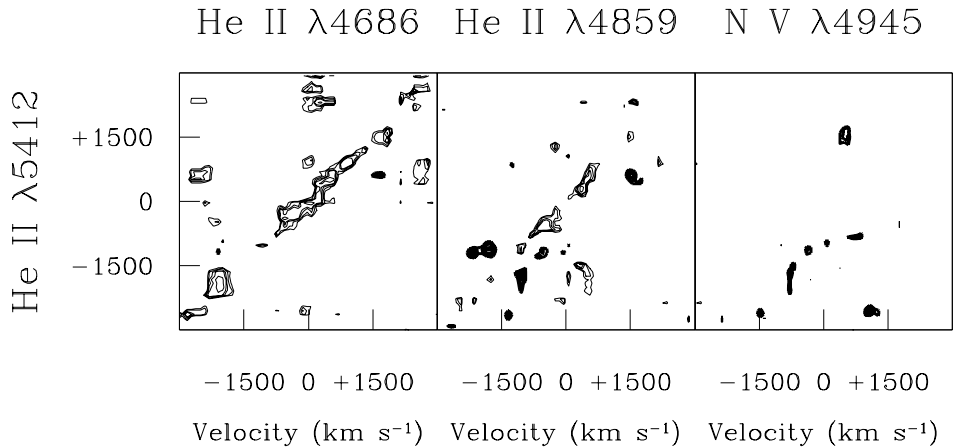}}
\caption{Correlation matrices of \ion{He}{ii} $\lambda$5412 with \ion{He}{ii} $\lambda$4686, \ion{He}{ii} $\lambda$4859, and \ion{N}{v} $\lambda$4945.  Thick and thin contours indicate a negative
or positive correlation in the pattern of variability presented by two
 line profiles at different projected velocities (referred to the line laboratory rest wavelength), respectively. The
lowest contour is drawn for a significant correlation at the 99.5 \% confidence level.}
\label{f7}
\end{figure*}

A possible correlated pattern of variability in two different spectral features has been investigated by calculating the Spearman rank-order correlation
matrices (see, e.g., \cite{Johnsbasri}), whose elements $r(i,j)$ yield the degree of correlation between the line
intensity variations at pixels $i$ and $j$ in each line profile, respectively. In the case of perfectly, positively correlated variations, a matrix unity is obtained.
The correlation matrices of \ion{He}{ii} $\lambda$5412 with \ion{He}{ii} $\lambda$4686, \ion{He}{ii} $\lambda$4859, and \ion{N}{v} $\lambda$4945 are shown in Figure \ref{f7} in the form of contour
plots, where the
lowest contour indicates a significant positive (or negative) correlation at the 99.5 \%
confidence level (the other spectral lines were not covered enough to be included in the analysis). These matrices are displayed in the projected velocity frame
(referred to the line laboratory rest wavelength). A significant positive correlation is generally found between the pattern of variability of \ion{He}{ii} $\lambda$5412 and \ion{He}{ii} $\lambda$4686. The same conclusion, however restricted to the velocity range (-- 1000, + 1000) km s$^{-1}$, holds for the variations affecting \ion{He}{ii} $\lambda$5412 and \ion{He}{ii} $\lambda$4859. In contrast, the (weak) variations of \ion{N}{v} $\lambda$4945 are apparently not linked to those of \ion{He}{ii} $\lambda$5412 (the same is true for \ion{C}{iv} $\lambda$5806 and \ion{He}{i} $\lambda$5876). Various factors are susceptible to mask a potentially correlated pattern (e.g., noise, blends). Also, due to the stratified nature of WR winds, it is conceivable that (a) two given features present similar, yet time-delayed patterns of variability, and (b) the variability is mainly restricted to the line formation regions of the \ion{He}{ii} ions and does not extend where the bulk of the \ion{N}{v} $\lambda$4945 emission originates. This would also result in an apparent lack of correlation. Overall, the results of this analysis tend to show that the \ion{He}{ii} features vary in a fairly similar fashion. The small number of spectral features included in the analysis prevents, however, to draw at this stage more general conclusions.
\subsection{Centroid, FWHM, EW, and Skewness Variations}
\begin{figure*}
\resizebox{\hsize}{!}{\includegraphics{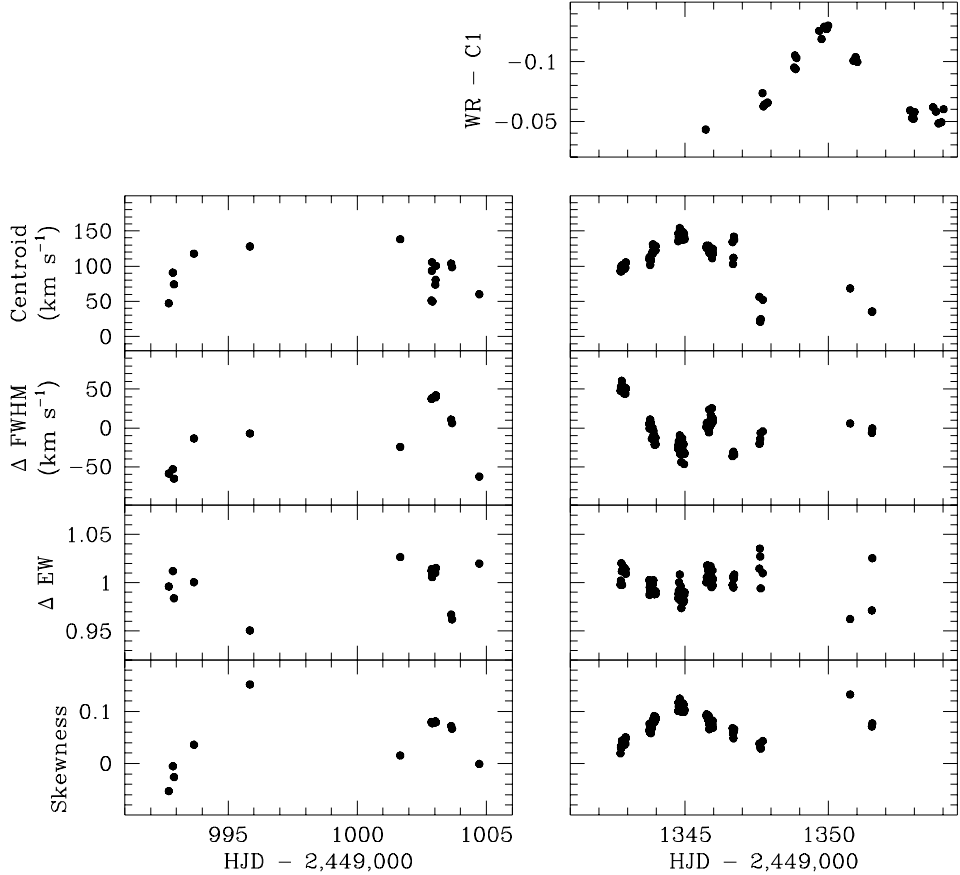}}
\caption{From top to bottom: continuum flux level variations, centroid
variations (in km s$^{-1}$), deviations
of the FWHM around the mean value (in km s$^{-1}$), EW variations (normalized by
division to
the mean value), and
skewness variations of \ion{He}{ii} $\lambda$4686 as a function of the
heliocentric Julian date of observation. \emph{Left panel}: 1995 October; \emph{right panel}: 1996 September.}
\label{f8}
\end{figure*}

We have investigated the time-dependent changes in the global line-profile
properties by calculating the centroid, FWHM,
EW, and skewness of \ion{He}{ii} $\lambda$4686, the strongest line. The centroid and skewness were calculated
as the first moment, and the ratio of the third and the (3/2
power of the) second central
moments of the portion of the profile above two in units of the
continuum (in order to avoid the contribution of blends). The FWHM was determined by a Gaussian
fit to the entire line profile. The EWs were calculated by
integrating the line flux in the interval 4643--4780 \AA. One can note that the mean EW of \ion{He}{ii} $\lambda$4686 increased from 313 to 343 \AA \ from 1995 October to 1996 September. This 10 \% increase may be explained by an intrinsic EW variability and/or by long-term changes in the stellar continuum flux. As can be seen
in Figure \ref{f8} (\emph{right panel}), the data obtained in 1996 September generally show \emph{coherent} time-dependent
variations. This is especially clear for the skewness time series, with
the timescale of the variations being of the same order of magnitude as the variations of
the continuum flux shown in the uppermost part of
Figure \ref{f8}, namely, in the range 5--7 days (note that, unlike the EW, the skewness is insensitive to changes in the continuum flux level). On the other hand,
however, and although this may be induced by the paucity of the data, no
clear time-dependent behavior is noticeable in the 1995 October dataset
(\emph{left panel}).

\section{Comparison with Previous Studies}
Niedzielski (1996a) reported on a spectroscopic pattern of variability of \object{WR 1} very much reminiscent of the one
presented by the apparently single WN 5 star \object{WR 6} that displays phase-locked (although strongly epoch-dependent) spectral variations according to $\cal P$ $\approx$ 3.77
days (\cite{Morel98a}, and references therein). The results of our investigation of the spectral variability of \object{WR 1} are
broadly consistent with this suggestion. In particular, as general features of the spectroscopic pattern of variability,
one can note in both objects the substantial variations of the absorption component
of the optical P Cygni profiles (readily observable in \ion{N}{v} $\lambda$$\lambda$4604, 4620 where the
absorption trough occasionally disappears; Fig.\ref{f5}) or the coherent time-dependent
changes in the global line-profile properties (e.g., skewness; Fig.\ref{f8}). 
Similarities can also be found with \object{WR 134}, another very rare example of single-line WR star displaying
cyclical variations (\cite{Morel98b}).

Evidence was also presented by Niedzielski (1996a) for cyclical (according to ${\cal P}$ $\approx$ 2.667 days), correlated changes in the EWs of \ion{He}{ii} $\lambda$4686
and  \ion{He}{ii} $\lambda$5412. The data presented in the present paper do not, however, support a claim of such periodicity. In particular, a period search in our EW data (but also in our centroid, FWHM, and skewness data; Fig.\ref{f8}) yields no significant signal at the expected frequency. Also, this 2.667 day period can hardly account for our global light-curve morphology (Fig.\ref{f1}). In a speculative vein, this apparent disagreement between our results and those presented by Niedzielski (1996a) \emph{might} point to a multiperiodic nature of the variability in \object{WR 1}, with the 2.667 day period being present in 1994-1995 but being dominated by other longer (non)cyclical processes during our observations. It has to be noted that based on a more comprehensive analysis of the data, Niedzielski (private communication) recently questioned the strictly periodic nature of the spectral variations. 

  As already noted by Niedzielski (1995, 1996b)
and  \cite{Wessoniedp}, our data provide evidence for a more gradual spectral pattern of variability than for \object{WR 6}; a fact which may imply that the period \emph{(if any)} is longer than 3.77 days. In support of this view, the centroid, EW, and skewness data of \ion{He}{ii}
$\lambda$4686 show coherent time-dependent patterns of variability (Fig.\ref{f8}), with
timescales in the range 5--6 days after a formal
periodicity search. However, we caution the reader that this period should not be taken too literally because of the limited time sampling of these data. In particular, a cyclical pattern
in this
range, as proposed by \cite{Lamontagnep} and \cite{Moffatp}, is clearly inconsistent with the lack of continuum flux variations observed after HJD 2,450,353 (Fig.\ref{f1}) unless one invokes a sudden period of ``quiescence'' after this date, with the small amplitude of the variations masking
any periodic patterns.
  
\section{On the Presence of a Companion}

Although a cyclical pattern of variability has yet to be unambiguously established
in \object{WR 1}, general considerations are made below regarding the model assuming an orbiting
(unseen) companion as
the origin of the variability. 
\subsection{A Non-degenerate Companion?}

\begin{figure}
\resizebox{\hsize}{!}{\includegraphics{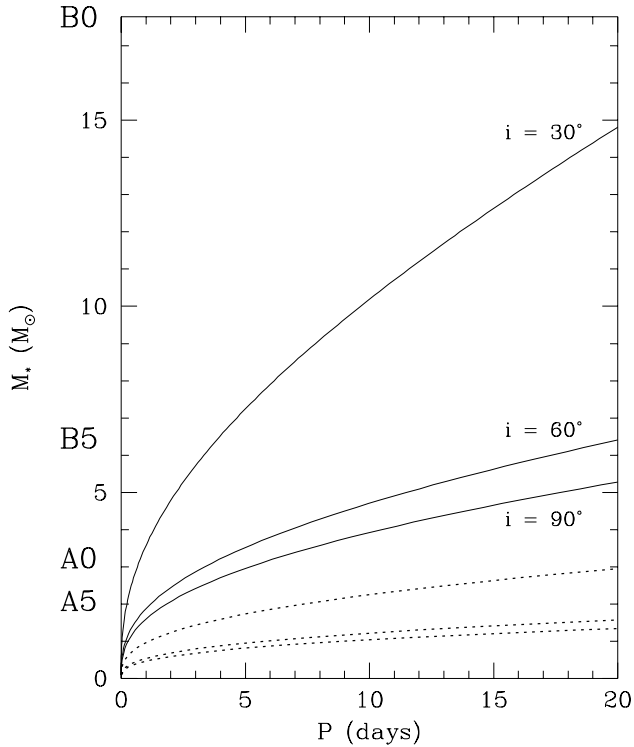}}
\caption{Allowed values of ($\cal P$, $M_{\star}$) when adopting $K_{WR}$ $\approx$ 70 km s$^{-1}$ (\emph{solid lines}) and $K_{WR}$ $\approx$ 22 km s$^{-1}$ (\emph{short-dashed lines}). These values are shown for different orbital inclinations: 30$^{\degr}$, 60$^{\degr}$, and 90$^{\degr}$. The spectral type of the companion (assuming a main-sequence type star) is indicated in the leftmost part of this figure.}
\label{f9}
\end{figure}

Some constraints can be set on the mass, $M_{\star}$, of this putative companion on the basis of the centroid measurements of \ion{He}{ii} $\lambda$4686 presented in Figure \ref{f8}. Assuming that these variations are \emph{entirely} attributable to orbital motion (i.e., $K_{WR}$ $\approx$ 70 km s$^{-1}$), one can explore the allowed values of ($\cal P$, $M_{\star}$). The solid lines in Figure \ref{f9} show the result of this investigation for three illustrative values of the orbital inclination ($i$ = 30$^{\degr}$, 60$^{\degr}$, and 90$^{\degr}$). In these calculations, we assume a circular orbit and a mass for \object{WR 1} of 9.1 M$_{\sun}$ (\cite{Hamann}). For a wide range in orbital inclination ($i$ $\ga$ 30$^{\degr}$) and in orbital period ($\cal P$ $\la$ 20 days), an upper limit for the companion's mass of 15 M$_{\sun}$ is derived. Assuming the companion to be a main sequence star, this constrains the spectral type to be later than B1. In this case, the companion's wind is too weak (\cite{Grigsby}) to induce wind-wind collision effects that may induce the large spectral changes observed in \object{WR 1}. For a system observed nearly face-on ($i$ $\la$ 30$^{\degr}$), larger masses are evidently consistent with the $K_{WR}$ value adopted above. However, no direct evidences (i.e., photospheric lines in the integrated spectrum or dilution of the WR continuum) support the presence of a luminous, early-type companion. In the presence of a non-degenerate companion, one would also expect a flat or eclipsing-type light curve. Yet, the \emph{opposite} behavior is observed in Figure \ref{f1}. These arguments strongly argue against the presence of a non-degenerate star orbiting \object{WR 1}.
\subsection{A Collapsed Companion?}

For a canonical mass of the companion as a neutron star, $M_{\star}$ $\approx$ 1.4 M$_{\sun}$, improbably small periods below one day are consistent with the adopted $K_{WR}$ value (Fig.\ref{f9}). However, it has to be kept in mind that the centroid measurements of \ion{He}{ii} $\lambda$4686 included the highly variable uppermost part of the profile which is unlikely to purely reflect orbital motion. Therefore, this $K_{WR}$ value is probably grossly overestimated. If one considers lower $K_{WR}$ values (as Lamontagne 1983: $K_{WR}$ = 22 $\pm$ 5 km s$^{-1}$; the allowed values of [$\cal P$, $M_{\star}$] are shown in this case by short-dashed lines in Fig.\ref{f9}), much larger periods (as in fact suggested by our study) are allowed. Thus, these considerations are not sufficient by themselves to rule out the presence of a neutron star companion (a black hole companion, on the other hand, would require long periods and/or small orbital inclinations). \footnote{Considering a lower $K_{WR}$ value in \S5.1 does not qualitatively modify our conclusions.}

In the presence of a collapsed companion, one may expect fairly strong,
accretion-type X-ray emission. Earliest observations by the \emph{HEAO
A--1} experiment gave an upper limit $L_X$ $\approx$ 4.4 $\times$ 
10$^{33}$ erg s$^{-1}$ on the emission in the 0.5--20 keV range (\cite{Helfand}). A value $L_X$ = 7.07 $\pm$ 2.85 $\times$ 10$^{32}$ erg s$^{-1}$ in the
0.2--2.4 keV range has been reported for \object{WR 1} during the
\emph{ROSAT} all-sky survey (\cite{Pollock}). \object{WR 1} is a fairly
strong X-ray emitter compared to
other (apparently) single WN stars. However, its
emission is by no means unusual when only considering the WNE-s subclass
(\cite{Wessolowski96}). Two subsequent pointed \emph{ROSAT} 
PSPC observations showed that a satisfactory fit to the X-ray
spectrum can be achieved, either with  a
\emph{Raymond-Smith} thermal plasma of about 1 keV or by the model
developed by \cite{Baum}, assuming a mixture of ``cool'' (in radiative
equilibrium) and ``hot'' (shocked) material (\cite{Wessolowski95};
\cite{Wessolowski96}). This picture is consistent with our current understanding of the
X-ray production in bona fide single WR stars, as being due to radiatively-induced instabilities (e.g., \cite{Willis}). This ``normal''
level of X-ray emission from \object{WR 1} does not
constitute, however, a decisive argument against the presence of a
collapsed companion, as the accretion of the 
wind material onto the neutron star is known to be inhibited in some X-ray binaries (e.g., \cite{Zhang}). 

Because of the spiral-in process that massive close
binaries are believed to experience in the course of their evolution, one is led to
expect periods of some hours for systems made up of a WR star and a compact companion, not
days (e.g., \cite{DeDonder}). A period of 4.8 hr is observed in 
\object{Cygnus X--3}, the prime candidate for a WR + compact companion system (\cite{vanKerkwijk}). 
\section{Concluding Remarks}
The single-star hypothesis appears appealing when considering the similarities
between the optical spectral pattern of variability of \object{WR 1} and the ones of the
two peculiar stars \object{WR 6} and \object{WR 134}. In this respect, although the light-curve morphology of WR
1 (a ``bump'' followed by a plateau) has, to our knowledge, no
example among the WR population (Moffat \& Shara 1986; \cite{Lamontmoffat}; \cite{Antokhin95}; \cite{Marchenko98b}), such a well-defined
light-curve pattern is reminiscent of what is observed in \object{WR 6}
(\cite{Robertetal}).\footnote{Curiously, a very similar light-curve morphology (both
in terms of the timescales involved and of the amplitude of the
variations) has been noticed in the Be star \object{FV CMa} (\cite{Balona}).} Such a repeatable pattern would, however, imply an unlikely large value for the period in the context of this single-star
hypothesis ($\cal P$ $\ga$ 18 days). 

Since the substantial
depolarization of the emission lines observed in \object{WR 6} and \object{WR 134} is generally taken as evidence for an equatorially-enhanced outflow (\cite{SchulteLadbeck91}, 1992), revealing the same phenomenon in \object{WR 1} may lead to the interesting suggestion that the occurence of large-scale line-profile and photometric variations (and thus possibly of azimuthally structured outflows) in single WR stars is somehow linked to the existence of a wind-compressed zone (\cite{Ignace}).

\begin{acknowledgements}
We acknowledge an anonymous referee and Alex W. Fullerton, whose comments have stimulated a substantial improvement of this manuscript.      T. M., Y. G., and N. S.-L. wish to thank the Natural Sciences
 and Engineering Research Council (NSERC) of Canada and the Fonds
pour la Formation de Chercheurs et l'Aide \`a la Recherche (FCAR) of Qu\'ebec for financial
 support; T. E. is grateful for full financial aid from the
Evangelisches Studienwerk/Germany, which is supported by the German
Government.
\end{acknowledgements}

\end{document}